\documentstyle[sprocl]{article}

\def\beq{\begin{equation}}
\def\eeq{\end{equation}}

\def\beqn{\begin{eqnarray}}
\def\eeqn{\end{eqnarray}}
\def\nn{\nonumber\\}

\def\O{{\mathcal O}}
\def\Im{\,\mbox{Im}\,}

\begin{document}

\title{SPIN POLARIZABILITIES OF THE NUCLEON}
\author{A. I. L'VOV}
\address{P. N. Lebedev Physical Institute, Russian Academy of Sciences
  \\ Leninsky Prospect 53, Moscow 117924, Russia }
\maketitle

\abstracts{
Spin polarizabilities of the nucleon ($\gamma_i$) are discussed in the
framework of fixed-$t$ and backward-angle dispersion relations, chiral
perturbation theory, and the non-relativistic quark model.
Calculations with the dispersion relations generally confirm findings
from HBChPT and disagree with a recent experimental result from LEGS
for the backward spin polarizability.}

\section{Introduction}

With recent progress in developing and using effective field theories,
a practical knowledge of various low-energy parameters of hadrons and
their interactions becomes of high current interest. Among such
parameters are the dipole electric and magnetic polarizabilities of the
nucleon, $\alpha_E$ and $\beta_M$, and the so-called spin (or vector)
polarizabilities $\gamma_{E1}$, $\gamma_{M1}$, $\gamma_{E2}$,
$\gamma_{M2}$. They characterize a low-energy behavior of the nucleon
Compton scattering amplitude up to order $\O(\omega^3)$ and correspond
to the following effective interaction of the nucleon's internal
degrees of freedom with the probing soft electromagnetic
field:~\cite{babu98}
\beqn
  \frac{1}{4\pi} H_{\rm eff} &=&
  - \frac12\alpha_E\vec E^2  - \frac12\beta_M\vec H^2
  - \frac12\gamma_{E1}\vec\sigma\cdot\vec E
       \times\frac{\partial\vec E}{\partial t}
  - \frac12\gamma_{M1}\vec\sigma\cdot\vec H
       \times\frac{\partial\vec H}{\partial t}
 \nn && {} \qquad
  + \gamma_{E2} E_{ij}\sigma_i H_j
  - \gamma_{M2} H_{ij}\sigma_i E_j    + \ldots
\eeqn
Here $E_{ij}= \frac12(\nabla_i E_j + \nabla_j E_i)$ and $H_{ij}=
\frac12(\nabla_i H_j + \nabla_j H_i)$ are the quadrupole strengths of
the electric $\vec E$ and magnetic $\vec H$ fields, and the omitted
terms are of higher order in the photon energy $\omega$.  In
particular, the low-energy $\gamma N$ scattering amplitude at forward
or backward scattering angles reads in terms of these polarizabilities:
\beqn
  f(0^\circ) &=& ~~~f_{\rm B}(0^\circ) +
    \omega^2 (\alpha_E+\beta_M)\,\vec\epsilon^{\,\prime}\cdot\vec\epsilon
  + i\omega^3 \gamma ~ \,
    \vec\sigma\cdot\vec\epsilon^{\,\prime}\times\vec\epsilon
  + \ldots,
\\
  f(180^\circ) &=& f_{\rm B}(180^\circ) +
    \omega^2 (\alpha_E-\beta_M)\,\vec\epsilon^{\,\prime}\cdot\vec\epsilon
  + i\omega^3 \gamma_\pi\,
    \vec\sigma\cdot\vec\epsilon^{\,\prime}\times\vec\epsilon
  + \ldots,~~
\eeqn
where the Born term $f_{\rm B}= -(e^2 Z^2/4\pi M)\,
\vec\epsilon^{\,\prime}\cdot\vec\epsilon + \ldots$ is determined by the
electric charge $eZ$, the mass $M$, and the anomalous magnetic moment
of the nucleon, and the quantities
\beqn
\label{gamma-0}
   \gamma &=& -\gamma_{E1} -\gamma_{M1} -\gamma_{E2} -\gamma_{M2},
\\
\label{gamma-pi}
   \gamma_\pi &=& -\gamma_{E1} +\gamma_{M1} +\gamma_{E2} -\gamma_{M2}
\eeqn
are the so-called forward and backward spin polarizabilities.

For the proton, the dipole polarizabilities $\alpha_E$ and $\beta_M$
are reasonably well determined by experiments on low-energy $\gamma p$
scattering done in Moscow, Urbana-Champaign, Mainz, and Saskatoon
(see~\cite{macg95} and references therein). All the spin
polarizabilities can, in principle, be determined in the
next-generation experiments on $\gamma p$ scattering with the
(circularly) polarized beam and target.  Since fixed-$t$ dispersion
relations predict~\cite{babu98,drec98} rather reliably three of the
four spin polarizabilities (the exception is $\gamma_\pi$), the
available data on {\em unpolarized} proton Compton scattering were used
to extract the remaining parameter $\gamma_\pi$ giving~\cite{tonn98}
\beq
\label{tonn98}
   \gamma_\pi = -27.1 \pm 3.4,
\eeq
where the units hereafter are $10^{-4}~{\rm fm}^4$ and all errors,
including systematic and model-dependent uncertainties, are summed up
in quadrature. This finding challenges the existing theories (including
HBChPT~\cite{hemm98} and dispersion relations~\cite{lvov98}), because
theoretically a much larger $\gamma_\pi \simeq -37 \mbox{~to~} -40$ is
anticipated.  In the following, we briefly review and compare arguments
of these theoretical frameworks and add another argument in their
support which is based on the non-relativistic quark model.

\section{Spin polarizabilities in HBChPT}

To leading non-vanishing order $\O(p^3)$, the structure-dependent
(non-Born) part of the Compton scattering amplitude is determined by
diagrams of the effective chiral Lagrangian with one-pion loop and with
the $t$-channel $\pi^0$-exchange.  No free parameters, except for the
standard set of them including the pion mass $m_\pi$, the pion decay
constant $f_\pi=92.4$ MeV, and the nucleon axial coupling $g_A=1.26$,
appear to this order.  Explicitly, the spin polarizabilities in the
chiral limit of $m_\pi \to 0$ are equal to~\cite{bern95}
\beq
\label{LO}
   \gamma_{E1} =-5X + X_a, \quad
   \gamma_{M1} = -X - X_a, \quad
   \gamma_{E2} =  X - X_a, \quad
   \gamma_{M2} =  X + X_a,
\eeq
where $X$ and $X_a$ represent the loop and $\pi^0$-exchange
contributions, $\gamma_i^{(\pi N)}$ and $\gamma_i^{(\pi^0)}$, which are
of second and first order in $g_A$, respectively:
\beq
\vspace*{-0.5ex}
   X=\frac{e^2 g_A^2}{384\pi^3 f_\pi^2 m_\pi^2}
       =1.11\times 10^{-4}~{\rm fm}^4,
\quad
   X_a=\frac{e^2 g_A \tau_3}{32\pi^3 f_\pi^2 m_\pi^2}
       = \pm 11.3\times 10^{-4}~{\rm fm}^4
\eeq
($X_a$ is positive for the proton, $\tau_3=+1$).

Actually, the leading-order approximation (\ref{LO}) represents
properties of the static nucleon with a polarizable perturbative pion
cloud. Quantitatively, such a simple picture does not work well and it
needs further corrections, such as the nucleon recoil and the
$\Delta$-isobar excitation, which formally appear in higher orders of
the chiral expansion. As quick means to include the phenomenologically
important $\Delta$ contribution, a modified expansion was proposed, in
which the $N\Delta$ mass splitting $\Delta=M_\Delta-M \simeq 2m_\pi$ is
counted as the value of order $\O(p)$ too (or, a small energy scale
$\epsilon=\O(m_\pi,\Delta)$).  Then, not only the $\Delta$-pole
contribution, but also contributions $\gamma_i^{(\pi\Delta)}$ of
1-loops with intermediate $\pi\Delta$ states have to be kept to the
same order.  Numerically, however, the $\pi\Delta$ loops contribute
little to the spin polarizabilities,~\cite{hemm98} so that a simplified
result of such an approach is reduced to adding the $\Delta$-pole
contribution to the dipole magnetic spin polarizability $\gamma_{M1}$:
\beq
\label{Delta}
   \gamma_{M1}^{(\Delta)} = \frac{\mu^2_{N\Delta}}{4\pi\Delta^2},
\eeq
where $\mu_{N\Delta}$ is the transition magnetic moment.  Depending on
the value used for $\mu_{N\Delta}$, Eq.~(\ref{Delta}) gives
between~\cite{hemm98} $+2.4$ and~\cite{babu98} $+4.0$.

The backward spin polarizability $\gamma_\pi = 4X - 4X_a +
\gamma_{M1}^{(\Delta)} + \gamma_\pi^{(\pi\Delta)}$ is clearly dominated
by the $\pi^0$ exchange which gives $\gamma_\pi^{(\pi^0)} = -4X_a
\simeq -45$ for the proton, whereas other terms give $\gamma_\pi^{(\pi
N)} + \gamma_\pi^{(\Delta)}+ \gamma_\pi^{(\pi\Delta)} \simeq +7$ to
$+9$, see Table~1.

\section{Dispersion relations}

A physical origin of the spin polarizabilities which emerges from
dispersion relations is very close to that of the HBChPT approach.
Generally, nucleon Compton scattering is described by six invariant
amplitudes $A_i(\omega,\theta)$ which are functions of $\omega$ and the
scattering angle $\theta$.  In the limit of $\omega\to 0$, the non-Born
parts $a_i$ of the amplitudes $A_{2,4,5,6}$ at zero energy determine
the spin polarizabilities which are linear combinations of the
constants $a_i$.~\cite{babu98} Three of these amplitudes ($A_{4,5,6}$)
satisfy unsubtracted dispersion relations at fixed $\theta=0$ which
give
\beq
\label{DR0}
   a_i = \frac{2}{\pi} \int_{\rm thr}^\infty
             \Im A_i(\omega,0^\circ) \,\frac{d\omega}{\omega},
   \quad i=4,5,6,
\eeq
and allow to determine through unitarity and photoproduction amplitudes
those combinations of spin polarizabilities which do not depend on
$a_2$.  They are
\beq
\label{gammas-good}
  \gamma_{E1}+\gamma_{M1}, \quad
  \gamma_{E2}+\gamma_{M2}, \quad
  \gamma_{E1}+\gamma_{E2}
\eeq
(and thus also $\gamma_{M1}+\gamma_{M2}$ and the forward spin
polarizability $\gamma$, Eq.~(\ref{gamma-0})).  A determination of the
last parameter $a_2$ which enters the polarizability $\gamma_\pi$,
Eq.~(\ref{gamma-pi}), is not reduced to the knowledge of
photoproduction only and needs also a knowledge of $t$-channel
exchanges, including $\pi^0$, $\eta$, $\eta'$. A dispersion relation at
the backward angle $\theta=180^\circ$, which includes both $s$- and
$t$-channel parts,
\beq
\label{backward}
   \gamma_\pi = \int_{\rm thr}^\infty   \sqrt{1+\frac{2\omega}{M}}
    \Big(1+\frac{\omega}{M}\Big)  \sum_n P_n \Big(
    \sigma_{3/2}^n(\omega) - \sigma_{1/2}^n(\omega) \Big)
    \frac{d\omega}{4\pi^2\omega^3} + \gamma_\pi^t,
\eeq
has recently been shown to provide a reliable way to calculate $a_2$
and $\gamma_\pi$ using the photoabsorption cross sections with the
total helicity 1/2 and 3/2 and with the relative parity $P_n$ of the
produced states $n$.~\cite{lvov98}

In the chiral limit, the unitarity gives the imaginary parts of the
amplitudes $A_i$ to order $\O(p^3)$ as a phase space $\O(p)$ times a
square of the pion photoproduction amplitude $\O(p)$, the latter being
given by tree diagrams of $\gamma N \to \pi N$ calculated with the
static nucleon.  Accordingly, the unsubtracted fixed-$t$ dispersion
relations (\ref{DR0}) give in the chiral limit
\beqn
\label{Ch-limit}
  && \hspace*{-2em}
   \gamma_i = C \int_0^1 G_i(v) \,v^2 dv,
   \quad C=\frac{e^2 g_A^2}{16\pi^3 m_\pi^2 f_\pi^2},
   \quad G_{E1}(v)=-\frac12 - \frac{1-v^2}{4v} \ln\frac{1+v}{1-v},
\nn &&
   G_{E2}(v)=G_{E1}(v) + 1, \quad G_{M1}(v)=G_{M2}(v) =0,
\eeqn
where $v$ is the pion velocity (cf.~\cite{lvov93}).
Eq.~(\ref{Ch-limit}) exactly reproduces the loop contributions
(\ref{LO}) of HBChPT to the ``good" polarizabilities
(\ref{gammas-good}).  For $\gamma_\pi$, however, Eq.~(\ref{Ch-limit})
does not work because the amplitude $A_2$, which gets also a
contribution from the $\pi^0$ exchange, does not obey Eq.~(\ref{DR0}).
In this case the backward dispersion relation (\ref{backward}) can be
used instead very efficiently. It gives in the chiral limit:
\beq
\label{Ch-limit-t}
   \gamma_\pi = C \int_0^1 G_\pi(v) \,v^2 dv + \gamma_\pi^t,
   \quad G_\pi(v)=\frac{1-v^2}{2v} \ln\frac{1+v}{1-v},
\eeq
where $\gamma_\pi^t=-4X_a$ is the contribution of the $t$-channel
$\pi^0$-exchange found with the WZW coupling, and again exactly matches
Eq.~(\ref{LO}).

With more realistic photoproduction amplitudes like those from the SAID
code,~\cite{SAID} and with additional $\eta$ and $\eta'$ exchanges, the
dispersion relations predict large deviations from
Eq.~(\ref{LO}).~\cite{babu98,drec98,lvov98}  A major part of these
deviations is caused, however, by the $\Delta$ resonance, so that the
obtained results for $\gamma$'s turn out to be not too far from
predictions of HBChPT with the $\Delta$-isobar explicitly included
through the $\epsilon$-expansion, see Table~1.

\section{Higher resonances in NQM}

For HBChPT to have a practical success, contributions of resonances,
which are normally treated as low-energy constants (or counter-terms),
have to be small. Therefore, the important question is what are
contributions of higher resonances (beyond $\Delta(1232)$) to the spin
polarizabilities. A non-relativistic quark model (NQM) can be used to
answer this question.

Considering the Hamiltonian of the NQM and doing a special gauge
transformation which makes the Born contribution manifestly
separated,~\cite{lvov92}  one can find the low-energy expansion of the
Compton scattering amplitude and obtain explicit formulas for all the
spin polarizabilities.~\cite{lvov99}  They involve dipole and
quadrupole matrix elements of the electromagnetic current.  The $M1\to
M1$ and $E2\to M1$ transitions are saturated by the $\Delta$-isobar
excitation and are given by the transition magnetic dipole
$\mu_{N\Delta}$ and the electric quadrupole $Q_{N\Delta}$ moments.  The
rest contributions can be evaluated using a closure over the
$1\hbar\omega$ shell which leads to many simplifications.  Then the
final result reads
\beq
\label{NQM}
   \gamma_{M1} = \frac{\mu^2_{N\Delta}}{4\pi\Delta^2}, \quad
   \gamma_{E2} = -\frac{\mu_{N\Delta} Q_{N\Delta}}{8\pi\Delta}, \quad
   \gamma_{E1} = -\gamma_{M2}
     = -\frac{e^2(1+\tau_3)}{72\pi m_q^2 \omega_q^2},
\eeq
where $m_q\simeq 340$ MeV is the quark mass and $\omega_q\simeq 500$
MeV is the oscillator frequency.  The term with $1/\omega_q^2$
represents a joint effect of a few nucleon resonances in the mass range
between 1520 and 1700 MeV and is absent for the neutron ($\tau_3=-1$).

In the present context, the only important thing is that all the
resonances lying beyond the $\Delta$ region have a small effect on the
spin polarizabilities, see Table~1. This finding supports the physical
conclusion inferred from dispersion calculations that the values of the
spin polarizabilities of the nucleon are mostly related with low-energy
excitations and long-range periphery of the nucleon (i.e.\  low-energy
$\pi N$ production and the $\pi^0$-exchange). This explains why HBChPT
to leading non-vanishing order gives already the results which are
rather close to those found through a much more detailed dynamical
input.

\section{Conclusions}

The presented arguments, both original and borrowed from other authors,
show that the spin polarizabilities $\gamma_i$ mostly depend on the
low-energy dynamics of the nucleon.  Different evaluations of the spin
polarizabilities agree with each other.  For the backward spin
polarizability $\gamma_\pi$, they suggest a much larger value than that
found in the recent LEGS experiment.~\cite{tonn98}  In view of the
large discrepancy between the fundamental predictions for $\gamma_\pi$
done on the basis of HBChPT and dispersion relations, it would be
important to test the experimental number (\ref{tonn98}) in a dedicated
experiment with polarized particles.

\section*{Acknowledgments}

I am pleased to thank organizers of the Conference for their invitation
and support.  Useful discussions with A.M.~Nathan and G.~Krein are very
appreciated.  This work was partially supported by the RFBR grant
98-02-16534.

\begin{table}[htb]
\caption{Spin polarizabilities of the proton in HBChPT and in the
dispersion theory. The columns $(\pi N)$, $(\Delta)$, $(\pi\Delta)$ and
$(\pi^0)$ give the HBChPT contributions~\protect\cite{hemm98} to order
$\O(p^3)$ (for $\Delta$, a ``large" $M1$ coupling was used and the $E2$
excitation was also included, see~\protect\cite{babu98}); $N^*$ is an
NQM estimate for higher resonances; DR($s$) are contributions of
photoproduction in the dispersion theory found with the SAID
multipoles; DR($t$) is the $t$-channel contribution for the backward
dispersion relation (includes $\pi^0$, $\eta$, $\eta'$).  For all but
the last line ($\gamma_\pi$), dispersion relations used are those at
fixed $\theta=0$, Eq.~(\ref{DR0}); for $\gamma_\pi$, it is
Eq.~(\ref{backward}).}
\medskip
\begin{center} $
\begin{array}{|l| |r|r|r|r|r| |r|r|}
\hline
  10^{-4}~{\rm fm}^4 &  (\pi N)   &  (\Delta)  &  (\pi\Delta)
   & N^*  &  (\pi^0)  &  \mbox{DR}(s) &  \mbox{DR}(t)
\\ \hline
 \gamma_{E1}+\gamma_{M1} & -6.7 &  4.0  &  0.45 & -0.4 &       & -0.7 &      \\
 \gamma_{E2}+\gamma_{M2} &  2.2 &  0.75 & -0.23 &  0.4 &       &  2.2 &      \\
 \gamma_{E1}+\gamma_{E2} & -4.4 &  0.75 &  0.21 & -0.4 &       & -1.5 &      \\
 \gamma_{M1}+\gamma_{M2} &    0 &  4.0  &     0 &  0.4 &       &  3.0 &      \\
 \gamma                  &  4.4 & -4.75 & -0.21 &  0   &       & -1.5 &      \\
\hline\hline
 \gamma_\pi              &  4.4 &  4.75 & -0.21 &  0   & -45.3 &  7.0 & -46.6\\
\hline
\end{array}
$ \end{center}
\end{table}

\end{document}